\documentclass[11pt]{article}

\advance\topmargin by -\headsep \hbadness5000 \emergencystretch=1cm
\everydisplay{\mathsurround 0pt} 

\RequirePackage[OT1]{fontenc}
\RequirePackage{amsthm,amsmath}
\RequirePackage[numbers]{natbib}
\RequirePackage[colorlinks,citecolor=blue,urlcolor=blue]{hyperref}
\usepackage{setspace}

\RequirePackage{array}
\RequirePackage{amssymb}
\RequirePackage{amsmath}
\RequirePackage{amsthm}
\RequirePackage{amsmath}
\RequirePackage{enumerate}
\RequirePackage{subfigure}
\RequirePackage{graphicx}
\newcommand{\n}{^{(n)}}

\newcommand{\R}{\mathbb R}

\newcommand{\N}{\mathbb N}

\newcommand{\Z}{\mathbb{Z}}

\newcommand{\varthetab}{{\pmb \vartheta}}

\newcommand{\Deltab}{{\pmb \Delta}}

\newcommand{\taub}{{\pmb \tau}}

\newcommand{\Gammab}{{\pmb \Gamma}}

\newcommand{\zerob}{{\pmb 0}}

\newcommand{\pr}{^{\prime}}

\newtheorem{lem}{Lemma}[section]
\newtheorem{Prop}{Proposition}[section]

\newtheorem{theor}{Theorem}[section]

\begin{document}

\title{Simple, asymptotically distribution-free, optimal tests for circular reflective symmetry about a known median direction
}

\author{Christophe {\sc Ley}\footnote{E-mail address: chrisley@ulb.ac.be; URL: http://homepages.ulb.ac.be/\~{}chrisley} \,   and Thomas {\sc Verdebout}\footnote{E-mail address: thomas.verdebout@univ-lille3.fr} \vspace{0.5cm}\\
Universit\'e Libre de Bruxelles and Universit\' e 
Lille Nord de France}
\date{}

\maketitle

%

\begin{abstract}
In this paper, we propose optimal tests for reflective circular symmetry about a fixed median direction. The distributions against which optimality is achieved are the so-called $k$-sine-skewed distributions of Umbach and Jammalamadaka~(2009). We first show that sequences of $k$-sine-skewed models are locally and asymptotically normal in the vicinity of reflective symmetry. Following the Le Cam methodology, we then construct optimal (in the maximin sense) parametric tests for reflective symmetry, which we render semi-parametric by a studentization argument. These asymptotically distribution-free tests happen to be uniformly optimal (under any reference density) and are moreover of a very simple and intuitive form. They furthermore exhibit nice small sample properties, as we show through a Monte Carlo simulation study. Our new tests also allow us to re-visit the famous red wood ants data set of Jander~(1957). We further show that one of the proposed parametric tests can as well serve as a test for uniformity against cardioid alternatives; this test coincides with the famous circular Rayleigh (1919) test for uniformity which is thus proved to be (also) optimal against cardioid alternatives. Moreover, our choice of $k$-sine-skewed alternatives, which are the circular analogues of the classical linear skew-symmetric distributions, permits us a Fisher singularity analysis \emph{\`a la} Hallin and Ley~(2012) with the result that only the prominent sine-skewed von Mises distribution suffers from these inferential drawbacks. Finally, we conclude the paper by discussing the unspecified location case.  
\end{abstract}
Keywords: Circular statistics, Fisher information singularity, Rayleigh test for uniformity, skewed distributions, tests for symmetry

\section{Introduction}\label{intro}
\setcounter{equation} {0}

Symmetry is a fundamental and ubiquitous structural assumption in statistics, underpinning most of the classical inferential methods, be it for univariate data on the real line or for circular data. Its acceptance generally simplifies the statistician's task, both in the elaboration of new theoretical tools and in the analysis of a given set of observations. For instance, the classical models for circular data, such as, e.g., the von Mises, cardioid, wrapped normal or wrapped Cauchy distributions (see Mardia and Jupp 2000, Section 3.5) are all symmetric about their unique mode (this form of symmetry on the circle is called \emph{reflective symmetry}). However, quoting Mardia~(1972, \mbox{p. 10}), ``symmetrical distributions on the circle are comparatively rare'', and recent years have shown an increasing interest in non-symmetric models (see, e.g., Umbach and Jammalamadaka~2009, Kato and Jones~2010, Abe and Pewsey~2011 or Jones and Pewsey~2012); therefore, it is all the more important to be able to test whether the hypothesis of symmetry holds or not, in order to know whether the classical, well explored, or rather the modern, less explored, models shall be used. Since circular distributions are encountered in several domains of scientific investigation, with particular emphasis on the analysis of (i) phases of periodic phenomena (physics, biology, etc.) and (ii) directions (animal movements as a response to some stimulus, pigeon homing, earth sciences, etc.), practical examples needing tests for circular symmetry about a specific axis/direction associated with the experimental set-up under consideration are all but rare.

On the real line, testing for symmetry about a fixed center (the median) is a classical yet timeless issue, and consequently there exists an abundance of such tests. Essentially, these tests can be distributed in two distinct categories. The first class contains nonparametric tests which are based on given characteristics of the null hypothesis of symmetry; famous examples are the Cram\'er-von Mises type tests of Rothman and Woodrofe~(1972) or Hill and Rao~(1977), the Kolmogorov-Smirnov type test of Butler~(1969), the runs test of McWilliams (1990) or its modified version by Modarres and Gastwirth~(1996), to cite but these. While being able to detect asymmetry against (usually) any type of skew alternative, such tests suffer from nonparametric rates of convergence, hence need in general a large number of observations in order to become powerful. Tests belonging to the second class are not universally consistent but instead rather focus on some favored alternatives, against which they achieve (semi)parametric consistency rates and sometimes even are (semi)parametrically optimal; examples of this category are provided in Kozubowski and Panorska~(2004), Cassart \emph{et al.}~(2008) or Ley and Paindaveine (2009). If one further considers shifts in location as ``skew'' alternatives, then the classical sign, Wilcoxon and van der Waerden tests (see H\'ajek and Sid\'ak~1967), which respectively achieve optimality under double-exponential, logistic and normal distributions, also belong to this second category.

In the circular case, the null hypothesis of symmetry is way less explored. There exist essentially three proposals for such tests:
\begin{itemize}
\item[$\bullet$] Schach~(1969) constructs locally optimal linear rank tests against rotation alternatives, the circular analogue of a linear shift alternative. His construction comprises the circular sign and Wilcoxon tests. 
\item[$\bullet$] Universally consistent tests from the linear setting have been adapted to the circular case (such as the celebrated runs tests, see Pewsey~2004).
\item[$\bullet$] A ``true'' test for circular symmetry has been studied in Pewsey~(2004) by having recourse to the second sine moment about the fixed median direction, a classical measure of circular skewness first proposed by Batschelet~(1965).
\end{itemize}
(One should not confuse the problem of testing for reflective symmetry treated here with that of testing for $l$-fold symmetry on the circle; this issue has been addressed in Jupp and Spurr~1983 (see also Mardia and Jupp~1999, page 146).) The scarcity of existing tests for circular symmetry might at first sight seem puzzling as one may be tempted to say that all tests for (linear) univariate symmetry should be adaptable to the circular setup (such as done for the runs tests), replacing the real line by the compact $[-\pi,\pi]$. However, this translation from one setup to the other is not so straightforward, due to several facts including that the points at $\pi$ and $-\pi$ coincide. Furthermore, replacing the circle by the compact set $[-\pi,\pi]$ requires the choice of an arbitrary ``reference point" on the circle which will play the role of the zero on the real line; this sounds unrealistic. As stated in Pewsey~(2004), when the observations are distributed on a large arc of the circle, it is likely that adapted tests suffer from a loss of power. It seems also very unlikely that optimal tests on the real line will keep their optimality features on the circle, as nothing \emph{a priori} ensures that they behave well against the (certainly complicated) wrapped versions of the univariate skew distributions they were designed for. Thus, except for rotation alternatives, the category of optimal tests for reflective symmetry appears to be empty.

In view of the emptiness of the second category of tests and the growing interest in skew circular distributions, our aim in the present paper is to fill in this gap by proposing tests for circular reflective symmetry about a fixed center that behave extremely well against a certain (general) type of skew alternatives. More precisely, we shall build locally and asymptotically optimal (in the maximin sense) tests for symmetry against \emph{$k$-sine-skewed alternatives} (Umbach and Jammalamadaka~2009, Abe and Pewsey~2011), a broad class of recently proposed skew circular distributions that has received an increasing interest over the past few years (see Section~\ref{sine} for a description). In a nutshell, these skew distributions are obtained by perturbation of a base symmetric distribution via a factor involving sines and a parameter to regulate skewness. Apart from the general interest in these skew circular distributions, the motivations for this choice are mainly twofold. First, they are the circular analogues of the most studied and most used skew distributions on the real line, namely the skew-symmetric distributions (see Azzalini and Capitano~2003 or Wang \emph{et al.}~Ê2004) inspired from the skew-normal distribution proposed in the seminal paper Azzalini~(1985). Second, as we shall see, the resulting test statistics are based on the trigonometric sine moments, hence provide this classical measure of circular skewness as well as the test of Pewsey~(2004) with  so far not known optimality properties. As nice by-products, our findings also enable us to (i) build optimal tests for uniformity against cardioid alternatives, and (ii) discuss Fisher singularity issues exactly as in the linear case.

The backbone of our approach is the Le Cam methodology which, although of linear nature, lends itself well for a transcription to circular settings (and even, with much more complications, to data living on unit hyperspheres in higher dimensions, see Ley \emph{et al.}~2013). In a first stage, we will obtain optimal parametric tests, and then, by means of studentization arguments, we shall turn them into semiparametric ones, valid under the entire null hypothesis of symmetry and optimal not only, as is usually the case, under the symmetric base distribution their parametric antecedents are based on, but \emph{uniformly} optimal under \emph{any} given symmetric base distribution. We will hence derive, as Schach~(1969), a family of fully efficient semiparametric tests which, in our case, are always optimal. For a given density, our tests will thus behave asymptotically as the likelihood ratio tests, but they clearly improve on the latter by their simplicity and the fact that, thanks to the Le Cam approach, one can derive explicit power expressions against sequences of contiguous skew alternatives.

The paper is organized as follows. In Section~\ref{sine}, we first describe the family of $k$-sine-skewed distributions, then establish their \emph{ULAN} property in the vicinity of symmetry, the crucial step in the Le Cam approach, and discuss some aspects of this property. In Section~\ref{known}, we construct our optimal tests for reflective symmetry about a known center and investigate their asymptotic properties.  The finite-sample performances of all our tests for reflective symmetry are evaluated and compared to existing tests in a large Monte Carlo simulation study, see Section~\ref{sec:simus}. Our tests are then applied on a famous real-data set in Section~\ref{real}. In Section~\ref{cardio}, we show how our findings also allow us to produce tests of uniformity against cardioid alternatives. The Fisher information singularity issue is tackled in Section~\ref{sec:sing}. We conclude the paper with final comments and an outlook on the case of non-specified median direction in Section~\ref{sec:fc}, and an Appendix collects the technical proofs.

\section{$k$-sine-skewed distributions and the ULAN property}\label{sine}

\subsection{$k$-sine-skewed densities}\label{sec:density}

As briefly depicted in the Introduction, $k$-sine-skewed distributions are obtained by perturbation of a base symmetric density. Define the collection
\begin{eqnarray*}
\mathcal{F}&:=&\bigg\{f_0:f_0(x)>0\,\mbox{a.e.}, f_0(x+2\pi k)=f_0(x)\forall k\in\Z, f_0(-x)=f_0(x),\\
&&\quad\quad f_0\,\mbox{unimodal at}\,0,\int_{-\pi}^\pi f_0(x)dx=1\bigg\}
\end{eqnarray*}
of unimodal reflectively symmetric (about the zero direction) circular densities. The most well-known representatives of this collection are the von Mises, cardioid or wrapped Cauchy distributions, with respective densities $f_{{\rm VM}_\kappa}(x):=\frac{1}{2\pi I_0(\kappa)}\exp(\kappa\cos(x))$ for $\kappa>0$ ($I_0$ stands for the modified Bessel function of the first kind and order zero), $f_{{\rm CA}_\ell}(x):=\frac{1}{2\pi}(1+\ell\cos(x))$ for $\ell\in(0,1)$, and $f_{{\rm WC}_\rho}(x):=\frac{1-\rho^2}{2\pi}\frac{1}{1+\rho^2-2\rho\cos(x)}$ for $\rho\in(0,1)$. A location parameter $\theta\in[-\pi,\pi]$ is readily introduced as center of symmetry, leading to densities $f(x-\theta), x\in[-\pi,\pi]$, with mode $\theta$. Inspired by the classical one-dimensional skewing method of Azzalini and Capitanio~(2003), Umbach and Jammalamadaka~(2009) have skewed such symmetric densities $f_0$ by turning them into
$$
2f_0(x-\theta)G(\omega(x-\theta)),\quad x\in[-\pi,\pi],
$$ 
where $G(x)=\int_{-\pi}^xg(y)dy$ is the cumulative distribution function (cdf) of some circular symmetric density $g$ and $\omega$ is a weighting function satisfying for all $x\in[-\pi,\pi]$ the three conditions $\omega(-x)=-\omega(x)$, $\omega(x+2\pi k)=\omega(x)\forall k\in\Z$, and $|\omega(x)|\leq\pi$. This construction being too general and for the sake of mathematical tractability, Umbach and Jammalamadaka have particularized their choice to $G(x)=(\pi+x)/(2\pi)$, the cdf of the uniform circular distribution, and $\omega(x)=\lambda\pi\sin(kx), k\in\N_0,$ with $\lambda\in(-1,1)$ playing the role of a skewness parameter. This finally yields what we call the \emph{$k$-sine-skewed densities}
\begin{equation}\label{densitygen}
x\mapsto f^k_{\theta,\lambda}(x):=f_0(x-\theta)(1+\lambda\sin(k(x-\theta))),\quad x\in[-\pi,\pi],
\end{equation}
with location parameter $\theta\in[-\pi,\pi]$ and skewness parameter $\lambda\in(-1,1)$. When $\lambda=0$, no perturbation occurs and we retrieve the base symmetric density, otherwise (\ref{densitygen}) is skewed to the left ($\lambda>0$) or to the right ($\lambda<0$). Further appealing properties of $k$-sine-skewed distributions are that $f^k_{\theta,\lambda}(\theta-x)=f^k_{\theta,-\lambda}(\theta+x)$, $f^k_{\theta,\lambda}(\theta)=f_0(0)$ independently of the value of $\lambda$, and the two endpoints, $f^k_{\theta,\lambda}(\theta-\pi)$ and $f^k_{\theta,\lambda}(\theta+\pi)$, coincide. However, for $k\geq 2$, $f^k_{\theta,\lambda}$ happens to be multimodal, whereas, for $k=1$, multimodality only rarely occurs. This explains why Abe and Pewsey~(2011) have restricted their attention to the study of the densities
\begin{equation}\label{density}
x\mapsto f_{\theta,\lambda}(x):=f_0(x-\theta)(1+\lambda\sin(x-\theta)),\quad x\in[-\pi,\pi],
\end{equation}
which they have called sine-skewed circular densities (hence our terminology $k$-sine-skewed densities for general $k$). Abe and Pewsey have shown in their paper under which conditions the densities (\ref{density}) happen to be multimodal. In the present paper, we establish all our theoretical results and propose tests for general $k$-sine-skewed distributions. Note that, when $f_0$ is the circular uniform density, then (\ref{density}) corresponds to the cardioid density $f_{{\rm CA}_\lambda}$ with mode at $\theta+\pi/2 (\mbox{mod}\, 2\pi)$, hence, in passing, we will as well construct and analyze an optimal test for uniformity against the cardioid distribution.

Sine-skewed (and $k$-sine-skewed) distributions lend themselves pretty well for modeling real-data phenomena. Aside from Abe and Pewsey~(2011) where this aspect is thoroughly described, these skew-circular distributions have been used, \emph{inter alia}, in the analysis of the $\mbox{CO}_2$ daily cycle in the low 
atmosphere at a rural site (P\'erez \emph{et al.}~2012) and of forest disturbance regimes (Abe \emph{et al.}~2012). This, combined with the motivations stated in the Introduction, makes $k$-sine-skewed distributions a natural choice as asymmetric alternatives in the construction of tests for circular reflective symmetry. 

\subsection{The ULAN property for $k$-sine-skewed densities}\label{sec:ULAN}

As explained in the Introduction, we shall have recourse to the Le Cam methodology in order to construct locally and asymptotically optimal tests for reflective symmetry against $k$-sine-skewed alternatives. For the sake of generality and in view of future research (see Section~\ref{sec:fc}), we here do not assume $\theta$ to be fixed. This of course contains the $\theta$-fixed case, which we need in this paper. Let $X_1,\ldots,X_n$ be \mbox{i.i.d.} circular observations with common density~(\ref{densitygen}) (that is, the $X_i$'s are angles). For any symmetric base density $f_0\in\mathcal{F}$ and any $k\in\N_0$, denote by ${\rm P}^{(n)}_{\varthetab;f_0,k}$, where $\varthetab:=(\theta,\lambda)\pr\in[-\pi,\pi]\times(-1,1)$, the joint distribution of the $n$-tuple $X_1,\ldots,X_n$. Since, for $\lambda=0$, the density $f^k_{\theta,\lambda}$ reduces to $f_0$ and hence does not depend on $k$, we drop the index $k$ and simply write ${\rm P}^{(n)}_{\varthetab;f_0}$ at $\varthetab=\varthetab_0:=(\theta,0)\pr$. Any pair $(f_0,k)$ induces the parametric location-skewness model
$$\mathcal{P}^{(n)}_{f_0,k}:=\left\{{\rm P}^{(n)}_{\varthetab;f_0,k}:\varthetab\in[-\pi,\pi]\times(-1,1)\right\},$$
whereas any $k\in\N_0$ induces the nonparametric location-skewness model $\mathcal{P}^{(n)}_k:=\cup_{f_0\in\mathcal{F}}\mathcal{P}^{(n)}_{f_0,k}$. 

The very first step of our construction of tests for symmetry about a fixed center consists in establishing the \emph{Uniform Local Asymptotic Normality (ULAN)} property, in the vicinity of symmetry (i.e., at $\lambda=0$), of the parametric model $\mathcal{P}^{(n)}_{f_0,k}$. This property of $k$-sine-skewed distributions happens to be interesting \emph{per se}, as it paves the way to numerous other applications of the Le Cam theory (such as, e.g., the construction of tests for symmetry about an unspecified center or of the celebrated one-step optimal estimators, see \mbox{e.g.} van der Vaart~2002). ULAN requires the following mild regularity condition on the base densities $f_0$.\vspace{0.3cm}

Assumption\,\,(A). The mapping $x\mapsto f_0(x)$ is a.e.-$\mathcal{C}^1$ over $[-\pi,\pi]$ (or equivalently over $\R$ by periodicity) with a.e.-derivative $\dot{f_0}$.\vspace{0.3cm}

Most classical reflectively symmetric densities satisfy this requirement. Note that the $\mathcal{C}^1$ condition over a compact combined with the fact that $f_0>0$ entails that, letting $\varphi_{f_0}=-\dot{f_0}/f_0$, the Fisher information quantity for location $I_{f_0}:=\int_{-\pi}^\pi\varphi^2_{f_0}(x)f_0(x)dx$ is finite. ULAN of the parametric model $\mathcal{P}^{(n)}_{f_0,k}$ with respect to $\varthetab=(\theta,\lambda)^\prime$, in the vicinity of symmetry, then takes the following form.

\begin{theor}\label{ULAN}
Let $f_0\in\mathcal{F}$ and $k\in\N_0$, and assume that Assumption (A) holds. Then, for any $\theta\in[-\pi,\pi]$, the parametric family of densities $\mathcal{P}^{(n)}_{f_0,k}$ is ULAN at $\varthetab_0=(\theta,0)^\prime$ with central sequence
\begin{eqnarray*}
\Deltab^{(n)}_{f_0,k}(\theta)
&:=&
\left( 
\begin{array}{c}
\Delta^{(n)}_{f_0,k;1}(\theta)\\[1mm]
\Delta^{(n)}_{k;2}(\theta)
\end{array}
\right)\\
&:=&
\frac{1}{\sqrt{n}}\, \sum_{i=1}^n 
\left( 
\begin{array}{c}
\varphi_{f_0}(X_i-\theta)\\
\sin(k(X_i-\theta))
\end{array}
\right),
\end{eqnarray*}
and corresponding Fisher information matrix 
$${\Gammab}_{f_0,k}:=\left(
\begin{array}{cc}
\Gamma_{f_0,k;11}&\Gamma_{f_0,k;12}\\
\Gamma_{f_0,k;12}&\Gamma_{f_0,k;22}
\end{array}\right),$$
where $\Gamma_{f_0,k;11}:=I_{f_0}$, $\Gamma_{f_0,k;12}:=-\int_{-\pi}^\pi\sin(kx)\dot{f_0}(x)dx$ and \linebreak $\Gamma_{f_0,k;22}:=\int_{-\pi}^\pi\sin^2(kx)f_0(x)dx$.
More precisely, for any $\theta^{(n)}=\theta+O(n^{-1/2})$ and for any bounded sequence $\taub^{(n)}=(\tau_1^{(n)},\tau_2^{(n)})'\in\R^2$ such that $n^{-1/2}\tau_2^{(n)}$ belongs to $(-1,1)$ and $\theta^{(n)}+n^{-1/2}\tau_1^{(n)}$ remains in $[-\pi,\pi]$  (with a slight abuse of notation since, e.g., $-\pi-\epsilon=\pi-\epsilon$ for $\epsilon>0$; what we mean is that the perturbation of the circular location parameter $\theta$ is such that it still lies on the unit circle), we have, letting \linebreak $\Lambda^{(n)}:=\log(d{\rm P}^{(n)}_{(\theta^{(n)}+n^{-1/2}\tau_1^{(n)},n^{-1/2}\tau_2^{(n)})';f_0,k}/d{\rm P}^{(n)}_{(\theta^{(n)},0)';f_0,k})$,
\begin{eqnarray}
\Lambda^{(n)}
&=&\taub^{(n)'}\Deltab^{(n)}_{f_0,k}(\theta^{(n)})-(1/2)\taub^{(n)'}{\Gammab}_{f_0,k}\taub^{(n)}+o_{\rm P}(1)\nonumber\\
&&\label{Taylor}
\end{eqnarray}
and $\Deltab^{(n)}_{f_0,k}(\theta^{(n)})\stackrel{\mathcal{L}}{\rightarrow}\mathcal{N}_2(\zerob,{\Gammab}_{f_0,k})$, both under ${\rm P}^{(n)}_{(\theta^{(n)},0)\pr;f_0,k}$ as $n\rightarrow\infty$.
\end{theor}

The proof is given in the Appendix. One easily sees that the Fisher information for skewness $\Gamma_{f_0,k;22}$, and hence the cross-information quantity $\Gamma_{f_0,k;12}$, is finite by bounding $\sin^2$ by $1$ under the integral sign. Note that the constant $k$ has no effect on the validity of Theorem~\ref{ULAN}. Note also that $\Delta^{(n)}_{k;2}(\theta)$ does not depend on $f_0$, a fact that will become of great interest in the sequel. With this ULAN property in hand, we are ready to derive our optimal tests for reflective symmetry about a fixed center $\theta$, as explained below in Section~\ref{sec:opt}. Moreover, since we do not fix $\theta$ in Theorem~\ref{ULAN}, our result also paves the way for deriving optimal tests for symmetry about an unknown center; see Section~\ref{sec:fc}.

We conclude the present section with a brief discussion on the minimal conditions required for having the ULAN property. Indeed, in view of the proof of Lemma~\ref{QMD} in the Appendix which is the main step to demonstrate Theorem~\ref{ULAN}, Assumption~(A) can be further weakened into
\vspace{0.3cm}

Assumption\,\,(A$^{\rm min}$). The mapping $x\mapsto f_0^{1/2}(x)$ is differentiable in quadratic mean over $[-\pi,\pi]$ (or equivalently over $\R$ by periodicity) with quadratic mean or weak derivative $(f_0^{1/2})^\prime(x)$ and, letting $\psi_{f_0}(x)=-2(f_0^{1/2})\pr(x)/f_0^{1/2}(x)$, the Fisher information quantity for location $J_{f_0}:=\int_{-\pi}^\pi\psi^2_{f_0}(x)f_0(x)dx$ is finite.\vspace{0.3cm}

Quadratic mean differentiability of $f_0^{1/2}$, a classical requirement in the Le Cam framework, means that $\int_{-\pi}^\pi(f_0^{1/2}(x-h)-f_0^{1/2}(x)-h\psi_{f_0}(x))^2dx=o(h^2)$ as $h\rightarrow0$, which corresponds exactly to the integral in expression~(\ref{half}) of the proof of Theorem \ref{ULAN} with $(f_0^{1/2})\pr$ instead of $\frac{1}{2}\dot{f_0}(x)/f_0^{1/2}(x)$ and hence is the minimal condition in order to have the ULAN property of the parametric model $\mathcal{P}^{(n)}_{f_0,k}$. Note that, under Assumption (A), these two derivatives of course coincide a.e., as well as $\psi_{f_0}$ and $\varphi_{f_0}$, and $J_{f_0}=I_{f_0}$ (as already mentioned above, the $\mathcal{C}^1$ condition ensures finiteness of $I_{f_0}$, while in the weaker Assumption (A$^{\rm min}$) one needs to ask that $J_{f_0}<\infty$).

\subsection{Constructing Le Cam optimal tests on basis of the ULAN property}\label{sec:opt}

The central idea of the Le Cam theory we are using here is the concept of convergence of statistical models (\emph{experiments} in the Le Cam vocabulary). Quoting Le Cam~(1960), ``the family of probability measures under study can be approximated very closely by a family of a simpler nature''. The ULAN property is an essential ingredient in this approximation, as it allows to deduce that (see Le Cam~1986 for details) our parametric location-skewness model $\mathcal{P}^{(n)}_{f_0,k}$ is locally (around $(\theta,0)\pr$) and asymptotically (for large sample sizes) equivalent to a simple Gaussian shift model. Intuitively, this follows from the fact that the likelihood ratio expansion~(\ref{Taylor}), if we neglect the remainder terms, strongly resembles the likelihood ratio of a Gaussian shift model $\mathcal{N}_2({\Gammab}_{f_0,k}\taub^{(n)},{\Gammab}_{f_0,k})$ with a single observation denoted by $\Deltab^{(n)}_{f_0,k}$. Since the optimal procedures for Gaussian shift experiments are well-known, we can translate them into our circular location-skewness model and hence obtain inferential procedures that happen to be asymptotically optimal (here, in the maximin sense). 

In the present paper, we shall employ this working scheme for testing the null hypothesis $\mathcal{H}_0^\theta$ of symmetry about a known central direction $\theta\in[-\pi,\pi]$. As explained above, our procedures will be (asymptotically) optimal against a fixed $k$-sine-skewed alternative~(\ref{densitygen}). For the considered testing problem, we first construct $f_0$-parametric tests for $\mathcal{H}_{0;f_0}^\theta={\rm P}^{(n)}_{(\theta,0)\pr;f_0}$: the optimality of these tests under the base density $f_0$ is thwarted by the fact that they are only valid (in the sense that they meet the asymptotic level-$\alpha$ constraint) under $f_0$. In order to palliate this non-validity outside $f_0$, we have recourse to a classical \emph{studentization argument} allowing us to turn our parametric tests  into tests for the semi- parametric null hypothesis $\mathcal{H}_{0}^\theta=\cup_{f_0\in\mathcal{F}}{\rm P}^{(n)}_{(\theta,0)\pr;f_0}$. The next section contains the detailed derivations of these tests.

\section{The test statistic and its asymptotic properties}\label{known}

Fix $\theta\in[-\pi,\pi]$. The $f_0$-parametric test $\phi^{(n);\theta}_{f_0;k}$ for circular reflective symmetry about a known central direction $\theta$ we propose rejects $\mathcal{H}_{0;f_0}^\theta$ at asymptotic level $\alpha$ whenever the statistic
\begin{equation}\label{parstat}
Q^{(n);\theta}_{f_0;k}:=\frac{|\Delta^{(n)}_{k;2}(\theta)|}{\Gamma^{1/2}_{f_0,k;22}}=\frac{|n^{-1/2}\sum_{i=1}^n\sin(k(X_i-\theta))|}{\Gamma^{1/2}_{f_0,k;22}}
\end{equation}
exceeds $z_{\alpha/2}$, the $\alpha/2$ upper quantile of the standard normal distribution (tests for reflective symmetry against one-sided alternatives of the form $\lambda>0$ or $\lambda<0$ are built similarly). It follows from the Le Cam theory that this test is locally and asymptotically maximin for testing the null $\mathcal{H}_{0;f_0}^\theta$ against  $\mathcal{H}_{1;f_0,k}^\theta:=\cup_{\lambda\neq0\in(-1,1)}{\rm P}^{(n)}_{(\theta,\lambda)\pr;f_0,k}$ (this optimality does not hold for $k'\neq k$).

Now consider $g_0\in\mathcal{F}$. Under ${\rm P}\n_{(\theta,0)\pr;g_0}$, $\Delta^{(n)}_{k;2}(\theta)$ is asymptotically normal with mean 0 and variance $\Gamma_{g_0,k;22}\neq\Gamma_{f_0,k;22}$. It is 
therefore natural to consider the \emph{studentized} test $\phi^{*(n);\theta}_{k}$ that rejects (at asymptotic level~$\alpha$) the null of circular reflective symmetry $\mathcal{H}_{0}^\theta$  as soon as 
\begin{equation}\label{studtest1}
Q^{*(n);\theta}_{k}:=\frac{|n^{-1/2}\sum_{i=1}^n\sin(k(X_i-\theta))|}{\left(n^{-1}\sum_{i=1}^n\sin^2(k(X_i-\theta))\right)^{1/2}}
\end{equation}
exceeds $z_{\alpha/2}$. We attract the reader's attention on the fact that this very simple test statistic does no more depend on $f_0$ (hence the omitment of the index $f_0$ in $\phi^{*(n);\theta}_{k}$). This is of course due to the fact that the central sequence for skewness, $\Delta^{(n)}_{k;2}(\theta)$, does \emph{not} depend on $f_0$. This remarkable fact implies that all parametric tests $\phi^{(n);\theta}_{f_0;k}, k\in\N,$ lead to the same studentized test statistic $\phi^{*(n);\theta}_{k}$, which therefore inherits optimality from its parametric antecedents under \emph{any} base symmetric distribution! This nice property as well as the asymptotics of such tests, under any $f_0\in\mathcal{F}$, follow from the ULAN property in Theorem~\ref{ULAN} and are summarized in the following result (see the Appendix for a proof).

\begin{theor} \label{theorstudent1}
Let $k\in\N_0$.  Then,

\noindent(i) under $\cup_{f_0\in\mathcal{F}}{\rm P}\n_{(\theta,0)\pr;f_0}$, $Q^{*(n);\theta}_{k}\stackrel{\mathcal{D}}{\rightarrow} \mathcal{N}(0,1)$ as $n\to\infty$, so that  the  test~$\phi^{*(n);\theta}_{k}$ has asymptotic level $\alpha$ under the same hypothesis;

\noindent(ii) under ${\rm P}^{(n)}_{(\theta,n^{-1/2}\tau\n_2)\pr;f_0,k'}$ with $f_0\in\mathcal{F}$ and $k'\in\N_0$, $Q^{*(n);\theta}_{k}$ is asymptotically normal with mean $\Gamma_{f_0,k;22}^{-1/2} C_{f_0}(k,k')\tau_2$ and variance $1$, where $\tau_2=\lim_{n\rightarrow\infty}\tau\n_2$ and $C_{f_0}(k,k'):=\int_{-\pi}^\pi \sin(kx)\sin(k' x)f_0(x)\,dx$ (which is finite);

\noindent(iii) 
for all $f_0\in\mathcal{F}$, $Q^{*(n);\theta}_{k}=Q^{(n);\theta}_{f_0;k}+o_{\rm P}(1)$ as $n\to\infty$ under ${\rm P}\n_{(\theta,0)\pr;f_0}$, so that the studentized test~$\phi^{*(n);\theta}_{k}$ is locally and asymptotically maximin, at asymptotic level~$\alpha$, when testing $\mathcal{H}_0^\theta$ against alternatives of the form $\cup_{\lambda\neq 0\in(-1,1)}\cup_{f_0\in\mathcal{F}}{\rm P}^{(n)}_{(\theta,\lambda)\pr;f_0,k}$.
\end{theor}
\vspace{2mm}

Theorem~\ref{theorstudent1}(i) shows that indeed the studentized test $\phi^{*(n);\theta}_{k}$ is valid under the entire null hypothesis $\mathcal{H}_{0}^\theta$, hence is asymptotically distribution-free. Note the uniform (in $f_0$, not in $k$) optimality of our studentized test. 

\begin{figure}
  \begin{center}
\includegraphics[width=8cm]{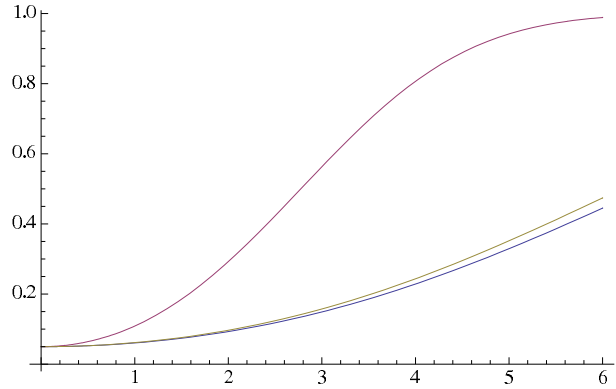}
  \end{center}
  \caption{Power curves, as a function of $\tau_2$, of the studentized test $\phi^{*(n);\theta}_{k}$ for $k=2$ against local alternatives ${\rm P}^{(n)}_{(\theta,n^{-1/2}\tau\n_2)\pr;f_0,k'}$ for $f_0$ the von Mises density with concentration parameter 1 and for $k'$ equal to 1 (blue line), 2 (red line) and 3 (yellow line).}
\label{figpo}
\end{figure}

For the sake of generality, we have also considered above alternatives where the natural $k$ is replaced by some $k'$ possibly different from the $k$ used in the construction of our tests. Point~(ii) allows us to give the explicit power of $\phi^{*(n);\theta}_{k}$ against the local alternatives ${\rm P}^{(n)}_{(\theta,n^{-1/2}\tau\n_2)\pr;f_0,k'}$:
$$
1-\Phi\left(z_{\alpha/2}-(\Gamma_{f_0,k;22})^{-1/2} C_{f_0}(k,k')\tau_2\right)+\Phi\left(-z_{\alpha/2}-(\Gamma_{f_0,k;22})^{-1/2} C_{f_0}(k,k')\tau_2\right),
$$
where $\Phi$ stands for the cdf of the standard Gaussian distribution. For the sake of illustration we have plotted, in Figure~\ref{figpo},  the power as a function of $\tau_2$ for $f_0$ the von Mises density with concentration parameter 1 and for $k=2$ and $k'=1,2,3$. Quite interestingly, when $k=2$, the statistic $\phi^{*(n);\theta}_{2}$ coincides exactly with the so-called ``b2-star'' test proposed in Pewsey~(2004). We have thus shown that this test enjoys maximin optimality features against $2$-sine-skewed alternatives, and provided its asymptotic powers against contiguous alternatives. This not only complements, but also gives further insight into the b2-star test. Finally, the very simple tests we have obtained are also easy to interpret as they are based on trigonometric sine moments, which are classical measures of skewness for circular data (see, e.g., Batschelet~1965).

\section{Monte Carlo simulation study}\label{sec:simus}

The aim of this section consists in examining the finite-sample performances of the proposed testing procedures for reflective symmetry (check of the nominal level constraint under distinct forms of reflective symmetry and investigation of power properties under various forms of asymmetry). To this end, we have generated $N=3,000$ independent samples of size $n=100$ from reflectively symmetric and increasingly skewed ($\lambda>0$) circular distributions, and run our tests as well as a natural competitor of the literature (the modified runs test of Pewsey~2004) under two-sided form at the asymptotic level $\alpha=5\%$. 

Let us fix the center of symmetry $\theta$ to $0$. We have performed our tests $\phi^{*(n);0}_1,\phi^{*(n);0}_2$ and $\phi^{*(n);0}_3$ as well as the modified runs test $\phi^{(n)}_{\rm modrun}$ with $p=0.6$. The latter test, initially proposed on the real line by Modarres and Gastwirth~(1996) as a modification of the classical runs test of McWilliams~(1990), has been adapted in Pewsey~(2004) to the circular case (our choice of $p=0.6$ also stems from that paper). The reason for choosing $\phi^{(n)}_{\rm modrun}$ as competitor lies in the simulation study driven in Pewsey~(2004). Indeed, recall that $\phi^{*(n);0}_2$ coincides with the so-called b2star test of \linebreak Pewsey~(2004); in that paper, the author has compared the performances of the test $\phi^{*(n);0}_2$ against various competitors, and the conclusion was that only $\phi^{(n)}_{\rm modrun}$ could compete with $\phi^{*(n);0}_2$, whence our restriction. Of course, we here choose other settings as in Pewsey~(2004), and hence also add new simulation-based information on Pewsey's b2star test. We also remark that we consider here $k=1,2,3$ for our tests because these values are able to capture both skew unimodality ($k=1$) and multimodality, but do not lead to too many oscillations of the sines within $[-\pi,\pi]$.

As reflectively symmetric distributions representing the null hypothesis, we have considered the von Mises laws $f_{{\rm VM}_1}$ and $f_{{\rm VM}_{10}}$, the cardioid $f_{{\rm CA}_{0.5}}$, the wrapped Cauchy $f_{{\rm WC}_{0.5}}$ as well as a mixture of two $f_{{\rm VM}_1}$ and two $f_{{\rm VM}_{10}}$ von Mises laws  with, in each case, respective centers at $-\pi/4$ and $\pi/4$ and mixing probability $0.5$. The latter mixture is used in order to assess the performances of our tests under bimodality. The densities $f_{{\rm VM}_1}$ and $f_{{\rm VM}_{10}}$ have then been turned into their $1$-,$2$- and $3$-sine-skewed versions, whereas $f_{{\rm CA}_{0.5}}$ and $f_{{\rm WC}_{0.5}}$ have become $1$- and $2$-sine-skewed. More concretely, in each case the skewness parameter $\lambda$ increases from zero to successively positive values. The bimodal mixture of von Mises laws has been skewed by simply shifting the center $\pi/4$ to $\pi/4+\lambda$. In order to also investigate other forms of perturbation of symmetry, we have applied the Moebius transform of Kato and Jones~(2012) to $f_{{\rm VM}_1}$ and $f_{{\rm VM}_{10}}$ with $r=0.5$ (in short, it turns $X$ into $\lambda+2\arctan\left(\omega_r\tan\left(\frac{X-\lambda}{2}\right)\right)$ with $\omega_r=\frac{1-r}{1+r}$). The rejection frequencies are reported in Table~\ref{Tab1} for $1$-sine-skewed alternatives, Table~\ref{Tab2} for $2$-sine-skewed alternatives and Table~\ref{Tab3} for Moebius, von Mises mixtures and $3$-sine-skewed alternatives.

\begin{table}
  \caption{Rejection frequencies (out of $N =3, 000$ replications), under various reflectively symmetric and $1$-sine-skewed distributions, of the optimal tests $\phi^{*(n);0}_1,\phi^{*(n);0}_2$ and $\phi^{*(n);0}_3$ as well as of the modified runs test $\phi^{(n)}_{\rm modrun}$ with $p=0.6$. The tests are performed at level $\alpha=5\%$.}
\footnotesize
\begin{center}{
\begin{tabular}{lcccccl}
 & \multicolumn{4}{c} {$1$-sine-skewed $f_{{\rm VM}_1}$}  \\[.5mm]
\hline
Test & $\lambda=0$ & $\lambda=0.2$ & $\lambda=0.4$ &$\lambda=0.6$   \\
\hline
$\phi^{*(n);0}_1$ & 0.0517 & 0.2633 & 0.7600 & 0.9873                \\
$\phi^{*(n);0}_2$ & 0.0567 & 0.0927 & 0.2277 & 0.4373             \\
$\phi^{*(n);0}_3$ & 0.0497 & 0.0537 & 0.0593 & 0.0713            \\
$\phi^{(n)}_{\rm modrun}$ &   0.0577 & 0.0747 & 0.1620   &0.3667           \\
\hline
 & \multicolumn{4}{c} {$1$-sine-skewed $f_{{\rm VM}_{10}}$}  \\[.5mm]
\hline
Test & $\lambda=0$ & $\lambda=0.2$ & $\lambda=0.4$ &$\lambda=0.6$   \\

\hline
$\phi^{*(n);0}_1$ &  0.0460 & 0.0993 & 0.2333 & 0.4573            \\
$\phi^{*(n);0}_2$ & 0.0470 & 0.0963 & 0.2303 & 0.4523           \\
$\phi^{*(n);0}_3$ & 0.0507  & 0.0900 & 0.2137 & 0.4203            \\
$\phi^{(n)}_{\rm modrun}$ &  0.0583 & 0.0610 & 0.0880 & 0.1387             \\
\hline
 & \multicolumn{4}{c} {$1$-sine-skewed $f_{{\rm CA}_{0.5}}$} \\[.5mm]
\hline
Test & $\lambda=0$ & $\lambda=0.2$ & $\lambda=0.4$ &$\lambda=0.6$   \\

\hline
$\phi^{*(n);0}_1$ & 0.0493 &  0.2993 & 0.8220 & 0.9960               \\
$\phi^{*(n);0}_2$ & 0.0497 & 0.0620 & 0.0940 & 0.1883              \\
$\phi^{*(n);0}_3$ &   0.0497  & 0.0447 & 0.0527 & 0.0460              \\
$\phi^{(n)}_{\rm modrun}$ & 0.0487 & 0.0630 & 0.1243 & 0.2680      \\

\hline
 & \multicolumn{4}{c} {$1$-sine-skewed $f_{{\rm WC}_{0.5}}$} \\[.5mm]
\hline
Test & $\lambda=0$ & $\lambda=0.2$ & $\lambda=0.4$ &$\lambda=0.6$  \\

\hline
$\phi^{*(n);0}_1$ &  0.0510 & 0.2417 & 0.6943 & 0.9703              \\
$\phi^{*(n);0}_2$ &    0.0507 & 0.0853 & 0.1973 & 0.3827                    \\
$\phi^{*(n);0}_3$ &  0.0513 & 0.0637 & 0.0860 & 0.1250              \\
$\phi^{(n)}_{\rm modrun}$ & 0.0523 & 0.0683 & 0.1493 & 0.3410             \\
\hline
  \end{tabular}
}\end{center}
\label{Tab1}
\end{table}
\

\begin{table}
  \caption{Rejection frequencies (out of $N =3, 000$ replications), under various reflectively symmetric and $2$-sine-skewed distributions, of the optimal tests $\phi^{*(n);0}_1,\phi^{*(n);0}_2$ and $\phi^{*(n);0}_3$ as well as of the modified runs test $\phi^{(n)}_{\rm modrun}$ with $p=0.6$. The tests are performed at level $\alpha=5\%$.}
\footnotesize
\begin{center}{
\begin{tabular}{lcccccl}
 &   \multicolumn{4}{c} {$2$-sine-skewed $f_{{\rm VM}_{1}}$} \\[.5mm]
\hline
 Test & $\lambda=0$ & $\lambda=0.2$ & $\lambda=0.4$ &$\lambda=0.6$  \\
\hline
  $\phi^{*(n);0}_1$  &     0.0460 & 0.1100 & 0.2423 & 0.4917               \\
 $\phi^{*(n);0}_2$ &  0.0473 & 0.3040 & 0.8287 & 0.9950              \\
$\phi^{*(n);0}_3$& 0.0550 & 0.0907 & 0.2427 & 0.4823                \\
$\phi^{(n)}_{\rm modrun}$& 0.0577 & 0.0677 & 0.1137 & 0.2487               \\
\hline
 &  \multicolumn{4}{c} {$2$-sine-skewed $f_{{\rm VM}_{10}}$} \\[.5mm]
\hline
Test & $\lambda=0$ & $\lambda=0.2$ & $\lambda=0.4$ &$\lambda=0.6$  \\

\hline
$\phi^{*(n);0}_1$& 0.0450  & 0.1813 & 0.5753 & 0.9133              \\
$\phi^{*(n);0}_2$& 0.0457 & 0.1833 & 0.5843 & 0.9157               \\
$\phi^{*(n);0}_3$& 0.0467 & 0.1800 & 0.5653 & 0.9013              \\
$\phi^{(n)}_{\rm modrun}$& 0.0533 & 0.0627 & 0.1730 & 0.4130              \\
\hline
 &   \multicolumn{4}{c} {$2$-sine-skewed $f_{{\rm CA}_{0.5}}$} \\[.5mm]
\hline
 Test & $\lambda=0$ & $\lambda=0.2$ & $\lambda=0.4$ &$\lambda=0.6$  \\

\hline
$\phi^{*(n);0}_1$& 0.0507 & 0.0633 & 0.1027 & 0.1840                \\
$\phi^{*(n);0}_2$& 0.0507 & 0.2717 & 0.8313 & 0.9917               \\
$\phi^{*(n);0}_3$& 0.0520 & 0.0700 & 0.1097 & 0.1787             \\
$\phi^{(n)}_{\rm modrun}$& 0.0553 & 0.0703 & 0.1350 & 0.2853              \\

\hline
& \multicolumn{4}{c} {$2$-sine-skewed $f_{{\rm WC}_{0.5}}$} \\[.5mm]
\hline
 Test & $\lambda=0$ & $\lambda=0.2$ & $\lambda=0.4$ &$\lambda=0.6$  \\

\hline
$\phi^{*(n);0}_1$& 0.0493 & 0.0920 & 0.2347 & 0.4543       \\
$\phi^{*(n);0}_2$& 0.0597 & 0.2710 & 0.7913 & 0.9913               \\
$\phi^{*(n);0}_3$& 0.0563 & 0.0937 & 0.2600 & 0.5230           \\
$\phi^{(n)}_{\rm modrun}$& 0.0480 & 0.0640 & 0.1367 & 0.2587         \\
\hline
  \end{tabular}
}\end{center}
\label{Tab2}
\end{table}

\begin{table}
  \caption{Rejection frequencies (out of $N =3, 000$ replications), under various reflectively symmetric and various skewed distributions, of the optimal tests $\phi^{*(n);0}_1,\phi^{*(n);0}_2$ and $\phi^{*(n);0}_3$ as well as of the modified runs test $\phi^{(n)}_{\rm modrun}$ with $p=0.6$. The tests are performed at level $\alpha=5\%$.}
\footnotesize
\begin{center}{
\begin{tabular}{lccccc|}
 & \multicolumn{4}{c} {Moebius transformed $f_{{\rm VM}_1}$} \\[.5mm]
\hline
Test & $\lambda=0$ & $\lambda=0.2/3$ & $\lambda=0.4/3$ &$\lambda=0.2$   \\

\hline
$\phi^{*(n);0}_1$ &  0.0447 & 0.1263 & 0.3437 & 0.6487                 \\
$\phi^{*(n);0}_2$ & 0.0480 & 0.1567 & 0.4603 & 0.7923            \\
$\phi^{*(n);0}_3$ &  0.0497 & 0.1637 & 0.4567 & 0.7720            \\
$\phi^{(n)}_{\rm modrun}$ &  0.0563 & 0.0570 & 0.0777 & 0.1133        \\
\hline
&  \multicolumn{4}{c} {Moebius transformed $f_{{\rm VM}_{10}}$} \\[.5mm]
\hline
 Test & $\lambda=0$ & $\lambda=0.02$ & $\lambda=0.04$ &$\lambda=0.06$  \\

\hline
$\phi^{*(n);0}_1$& 0.0510 & 0.2303 & 0.6553 & 0.9357           \\
$\phi^{*(n);0}_2$& 0.0517 & 0.2327 & 0.6563 & 0.9363                \\
$\phi^{*(n);0}_3$& 0.0520 & 0.2337 & 0.6613 & 0.9380         \\
$\phi^{(n)}_{\rm modrun}$& 0.0527 & 0.0850 & 0.2020 & 0.4623             \\
\hline
 & \multicolumn{4}{c} {Skewed $f_{{\rm VM}_1}$ mixtures}  \\[.5mm]
\hline
Test & $\lambda=0$ & $\lambda=0.4$ & $\lambda=0.8$ &$\lambda=1.2$   \\

\hline
$\phi^{*(n);0}_1$ & 0.0470 & 0.1007 & 0.1460 & 0.0980           \\
$\phi^{*(n);0}_2$ &  0.0457 & 0.0550 & 0.1240 & 0.2550               \\
$\phi^{*(n);0}_3$ &  0.0493 & 0.0470 & 0.0537 & 0.0543                   \\
$\phi^{(n)}_{\rm modrun}$ &  0.0453 & 0.0600 & 0.0787 & 0.0890             \\
\hline
&   \multicolumn{4}{c} {Skewed $f_{{\rm VM}_{10}}$ mixtures} \\[.5mm]
\hline
 Test & $\lambda=0$ & $\lambda=0.2$ & $\lambda=0.4$ &$\lambda=0.6$  \\

\hline
$\phi^{*(n);0}_1$& 0.0507 & 0.1303 & 0.2463 & 0.3560            \\
$\phi^{*(n);0}_2$& 0.0510 & 0.0693 & 0.3500 & 0.9573              \\
$\phi^{*(n);0}_3$& 0.0490 & 0.6607 & 1.0000 & 1.0000               \\
$\phi^{(n)}_{\rm modrun}$& 0.0553 & 0.2373 & 0.8027 & 0.9923          \\
\hline

 & \multicolumn{4}{c} {$3$-sine-skewed $f_{{\rm VM}_1}$}  \\[.5mm]
\hline
Test & $\lambda=0$ & $\lambda=0.2$ & $\lambda=0.4$ &$\lambda=0.6$   \\

\hline
$\phi^{*(n);0}_1$ &  0.0563 & 0.0450 & 0.0603 & 0.0787          \\
$\phi^{*(n);0}_2$ & 0.0483 & 0.0963 & 0.2420 & 0.4657             \\
$\phi^{*(n);0}_3$ & 0.0503 & 0.3027 & 0.8247 & 0.9920            \\
$\phi^{(n)}_{\rm modrun}$ &  0.0550 & 0.0657 & 0.1343 & 0.2840                     \\
\hline
&  \multicolumn{4}{c} {$3$-sine-skewed $f_{{\rm VM}_{10}}$} \\[.5mm]
\hline
 Test & $\lambda=0$ & $\lambda=0.2$ & $\lambda=0.4$ &$\lambda=0.6$  \\

\hline
$\phi^{*(n);0}_1$&0.0480 & 0.2297 & 0.6963 & 0.9687    \\
$\phi^{*(n);0}_2$& 0.0470 & 0.2410 & 0.7347 & 0.9783              \\
$\phi^{*(n);0}_3$& 0.0473 & 0.2520 & 0.7530 & 0.9827             \\
$\phi^{(n)}_{\rm modrun}$& 0.0557 & 0.0890 & 0.1947 & 0.5207              \\
\hline
  \end{tabular}
}\end{center}
\label{Tab3}
\end{table}

\

All four tests meet the $5\%$  nominal level constraint under each reflectively symmetric
density considered, even under bimodality, and seem to be unbiased.  Under $k$-sine-skewed alternatives, the theoretical optimality features of our tests $\phi^{*(n);0}_k$ are confirmed. Quite remarkably, when the observations are highly concentrated ($f_{{\rm VM}_{10}}$ case), the differences between the three tests vanish, whereas under certain $k'$-sine-skewed densities the test $\phi^{*(n);0}_k$ for $k\neq k'$ exhibits low powers (especially when combining the indices $1$ and $3$).  We moreover notice that all our tests are powerful under the Moebius transformed skew densities and even under skewed von Mises mixture distributions with high concentration parameter $\kappa$, which shows that the proposed tests not only act well under the laws they are designed for. As an overall summary, we see that our three tests generally outperform the modified runs test.

\section{A real-data application}\label{real}

\begin{figure}
  \begin{center}
\includegraphics[width=8cm]{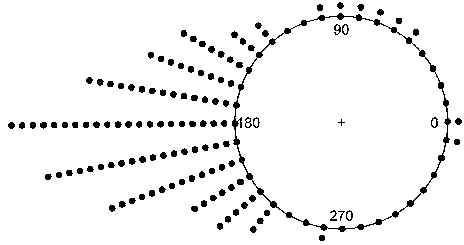}
  \end{center}
  \caption{Raw circular plot of the Jander~(1957) data set recorded during an orientation experiment with 730 red wood ants. Each dot represents the direction chosen by five ants.}
\label{figreal}
\end{figure}

In this section, we apply our optimal tests for reflective symmetry on a well-known data set in animal orientation problems. This data set stems from an experiment with 730 red wood ants (\emph{Formica rufa} L.) described in Jander~(1957). Each ant was individually placed in the center of an arena with a black target positioned at an angle of $180^\circ$ from the zero direction, and the initial direction in which each ant moved upon release was recorded to the nearest $10^\circ$. Thus it is clear that the experimental design suggests a natural median direction, a fact that is clearly corroborated by the graphical representation of the data in Figure~\ref{figreal}. The natural question of interest is whether the directions chosen by the ants are symmerically distributed around the median direction representing the black target. This data set has been discussed, \emph{inter alia}, in Pewsey~(2004), Umbach and Jammalamadaka~(2009) and Abe and Pewsey~(2011).

By the experimental design, this real-data set happens to be a very good candidate for testing circular symmetry about a known median direction (assuming the median direction unknown might even be inappropriate here and the corresponding tests for symmetry will not be as powerful as tests for symmetry about a fixed direction). In view of the data plot in Figure~\ref{figreal}, we see that the underlying density might be multimodal rather than unimodal, indicating that the tests $\phi^{*(n);0}_2$ and $\phi^{*(n);0}_3$ might be more powerful in the present situation than $\phi^{*(n);0}_1$ (we refer to Abe and Pewsey~Ê2011 for a discussion on the conditions under which $1$-sine-skewed distributions are unimodal or multimodal). Indeed, $\phi^{*(n);0}_1$ yields a p-value of 0.7781, while $\phi^{*(n);0}_2$ and $\phi^{*(n);0}_3$ respectively give p-values 0.0107 and 0.0131. This clearly provides evidence that the data are in fact not symmetrically distributed around the median direction of $180^\circ$. Pewsey~(2004) has obtained the same conclusion with his b2star test, but our conjunction of the tests $\phi^{*(n);0}_2$ and $\phi^{*(n);0}_3$ provides yet further information and evidence. In Abe and Pewsey~(2011), the authors notice that neither the symmetric nor the 1-sine-skewed distributions they have considered provide an adequate fit to this data. Their findings are not a surprise: according to $\phi^{*(n);0}_1$, 1-sine-skewed densities are not preferable over symmetric ones, while our other tests reject the hypothesis of reflective symmetry up to the $2\%$ level. This shows that, most probably, the ant data are best fitted by 2- or 3-sine-skewed distributions.

\section{Optimal test for uniformity against cardioid alternatives}\label{cardio}

In this section, we show that the parametric test statistic~(\ref{parstat}) not only serves for efficient testing of reflective symmetry but also for testing the null hypothesis $\mathcal{H}_0^{\rm unif}$ of uniformity against cardioid alternatives about a fixed central direction. Mardia and Jupp~(2000) explain that there exist circumstances under which one wishes to test uniformity against a certain fixed direction; such situations are the motivation behind our optimal test for uniformity. As described in Section~\ref{sec:density}, for $f_0(x)=\frac{1}{2\pi}$, the uniform density, (\ref{density}) with $k=1$ corresponds to the cardioid density with mode at $\theta+\pi/2\in[-\pi,\pi]$. Of course, we cannot directly use our ULAN result for this hypothesis, as the uniform distribution does not favour one particular direction, hence $\theta$ would not be identified under $\mathcal{H}_0^{\rm unif}$ (the uniform density does not belong to $\mathcal{F}$). However, writing ${\rm P}_{\rm unif}\n$ for the joint distribution of an $n$-tuple of uniform observations $X_1,\ldots,X_n$ and ${\rm P}_{(\theta, \lambda)'}\n$ for the joint distribution under the alternative $\mathcal{H}_1^{{\rm CA}_\lambda,\theta}$ of cardioid distribution, and using the Taylor expansion $\log(1+x)=x-\frac{x^2}{2} +o(x^2)$ around $0$, we obtain that
\begin{eqnarray*}
\log \left(\frac{d{\rm P}_{(\theta, n^{-1/2} \tau_2)'}\n}{d{\rm P}_{\rm unif}\n}\right) &=& \sum_{i=1}^{n} \log(1+n^{-1/2} \tau_2 \sin(X_i- \theta)) \\
&=& \tau_2  n^{-1/2} \sum_{i=1}^{n}\sin(X_i- \theta) - \frac{\tau_2^2}{2} {\rm E}[\sin^2(X_1- \theta)] +o_{\rm P}(1)
\end{eqnarray*}
as $n\rightarrow\infty$ under ${\rm P}_{\rm unif}\n$, which is nothing but a LAN decomposition. Hence, for fixed~$\theta$, an optimal test for testing $\mathcal{H}_0^{\rm unif}:\lambda=0$ against $\mathcal{H}_1^{{\rm CA}_\lambda, \theta}:\lambda\neq0$ shall be based on
\begin{equation}\label{uniftest}
Q^{(n)}_{{\rm unif};\theta}:=\sqrt{2}n^{-1/2} \sum_{i=1}^{n}\sin(X_i- \theta),
\end{equation}
that is, on~(\ref{parstat}) with $k=1$ and $\Gamma_{f_{\rm unif},k;22}=\frac{1}{2}$. The results of Theorem~\ref{theorstudent1} remain valid under this special parametric setup here (with, of course, the required modifications such as restricting to $f_0=f_{\rm unif}$ in Parts~(i) and~(iii) and restricting $k'$ to $1$ and $f_0$ to the cardioid density in Part~(ii)), and therefore provide the asymptotic behavior of $Q^{(n)}_{{\rm unif};\theta}$ and the working mechanism of the associated test of uniformity $\phi^{(n)}_{{\rm unif};\theta}$ (which is a one-sided test in view of the definition of the cardioid density). Now, recall that the central direction is not $\theta$ but $\theta+\pi/2$, hence the statistic~(\ref{uniftest}) should rather read
$$
Q^{(n)}_{{\rm unif};\theta}=\sqrt{2}n^{-1/2} \sum_{i=1}^{n}\sin(X_i- (\theta+\pi/2)+\pi/2)=\sqrt{2}n^{-1/2} \sum_{i=1}^{n}\cos(X_i- (\theta+\pi/2)),
$$
which corresponds exactly to the expression of the Rayleigh (1919) test of uniformity against a fixed central direction $\theta+\pi/2$, which is originally constructed to be efficient against von Mises alternatives, see Mardia and Jupp~(2000), pages 98 and 99. Thus, our findings now further indicate that the Rayleigh test is also optimal against cardioid alternatives. Note that  Theorem~\ref{theorstudent1} provides its asymptotic null and non-null behavior.

\section{Singularity of the location-skewness Fisher information matrix}\label{sec:sing}

Besides its numerous favourable properties, the skew-normal distribution of Azzalini~(1985) is also sadly famous for having a singular Fisher information matrix in the vicinity of symmetry, due to the collinearity of the scores for location and skewness. A vast literature has been devoted to the analysis of the reasons for this singularity, to possible cures and to the study of which other skew-symmetric distributions suffer from the same drawback. Indeed, a singular information matrix violates the assumptions for standard Gaussian asymptotics of the maximum likelihood estimators and precludes, at first sight, any nontrivial test of the null hypothesis of symmetry. Among the papers having dealt with this singularity issue are Azzalini~(1985), Azzalini and Capitanio~(1999), Pewsey~(2000), DiCiccio and Monti~(2004), Chiogna~(2005), Azzalini and Genton~(2008), Ley and Paindaveine~(2010) and Hallin and Ley~(2012). The latter paper exactly determines those skew-symmetric distributions that suffer from Fisher singularity. 

The present section can be inscribed into this stream of literature, as it discusses and solves the same problem for $k$-sine-skewed circular distributions. Moreover, our investigation shall become very important when one wants to construct optimal tests about an unknown center $\theta$, as will be briefly discussed in the final section. Now, recall that the information matrix in the vicinity of symmetry is given by
$${\Gammab}_{f_0,k}
=\left(
\begin{array}{cc}
\int_{-\pi}^\pi\varphi^2_{f_0}(x)f_0(x)dx&\int_{-\pi}^\pi\sin(kx)\varphi_{f_0}(x)f_0(x)dx\\
\int_{-\pi}^\pi\sin(kx)\varphi_{f_0}(x)f_0(x)dx&\int_{-\pi}^\pi\sin^2(kx)f_0(x)dx
\end{array}\right).$$
This matrix is singular if and only if 
\begin{equation}\label{sing}
\left(\int_{-\pi}^\pi\varphi^2_{f_0}(x)f_0(x)dx\right)\left(\int_{-\pi}^\pi\sin^2(kx)f_0(x)dx\right)=\left(\int_{-\pi}^\pi\sin(kx)\varphi_{f_0}(x)f_0(x)dx\right)^2.
\end{equation} 
The Cauchy-Schwarz inequality readily yields that the equality sign ``$=$'' in (\ref{sing}) can be replaced by ``$\geq$'' with equality holding if and only if $\varphi_{f_0}(x)=a\sin(kx)$ for some real constant $a$. The latter easy-to-solve first-order differential equation then shows that an information singularity can only occur for base symmetric densities $f_0$ of the form $c\exp(\frac{a}{k}\cos(kx))$ for $a\in\R$ and $c>0$ a normalizing constant. Now, bare in mind that the class of base densities $\mathcal{F}$ we consider contains the condition of unimodality on $f_0$, which directly rules out all values $k\geq2$ and forces $a$ to be positive. Hence, the only base symmetric density for which the Fisher information matrix ${\Gammab}_{f_0,k}$ is singular corresponds to $f_0(x)=c\exp(\kappa\cos(x))$ with $\kappa=\frac{a}{k}>0$ a concentration parameter, hence to the famous von Mises circular density. We formalize this result in the following proposition.
\begin{Prop}\label{singprop}
Let $f_0$ be a symmetric base density belonging to $\mathcal{F}$ and satisfying Assumption~(A), and consider $k$-sine-skewed densities of the form $f_0(x-\theta)(1+\lambda\sin(k(x-\theta)))$. Then the Fisher information matrix associated with the parameters $\theta\in[-\pi,\pi]$ and  $\lambda\in(-1,1)$ is singular in the vicinity of symmetry (that is, at $\lambda=0$) if and only if $k=1$ and $f_0(x)=c\exp(\kappa\cos(x))$ with $\kappa>0$ a concentration parameter and $c>0$ the normalizing constant, that is, if and only if one is considering sine-skewed von Mises densities.
\end{Prop}

Taking a closer look, this singularity only under the prominent von Mises densities is not really surprising. Indeed recall that the construction of asymmetric circular distributions by Umbach and Jammalamadaka~(2009) is inspired from the classical linear setup initiated by Azzalini~(1985).  For the univariate setting, it has been shown in Hallin and Ley~(2012) that the information matrix of skew-symmetric densities of the form $2f(x)G(\lambda x)$ (with self-explaining notations) is singular if and only if the base symmetric density $f$ is Gaussian. The circular equivalent of the linear first moment $x$ being $\sin(x)$ (since $\cos(x)$ is not odd), it is intuitively reasonable that the circular equivalent of the linear Gaussian density, that is, the von Mises distribution, suffers from Fisher singularity in the vicinity of symmetry, which allows to better understand the Fisher singularity result encountered here.

\section{Final comments}\label{sec:fc}

In this paper we have tackled the problem of testing circular reflective symmetry about a specified center. The tests we propose are uniformly (over the null hypothesis) locally and asymptotically maximin against $k$-sine-skewed alternatives, asymptotically distribution-free and moreover of a very simple form. They furthermore exhibit nice finite sample behaviors. Now, as already mentioned before, it would also be of interest to adapt our procedures to the case of an unspecified center, and our general ULAN property provides the required theoretical background for constructing such tests. The crucial difference to the tests of the present paper, of course, lies in the fact that we will need to replace the unknown location $\theta$ with an estimator $\hat\theta$. Would the information matrix ${\Gammab}_{f_0,k}$ be diagonal, then the substitution of $\hat{\theta}$ for $\theta$ would have no influence, asymptotically, on the behavior of the central sequence for skewness $\Delta^{(n)}_{k;2}(\theta)$. However, the covariance $\Gamma_{f_0,k;12}$ only rarely equals zero, hence a local perturbation of $\theta$ has the same asymptotic impact on $\Delta^{(n)}_{k;2}(\theta)$ as a local perturbation of $\lambda=0$. It follows that the cost of not knowing the actual value of the location $\theta$ is strictly positive when performing inference on $\lambda$; the stronger the correlation between $\theta$ and $\lambda$, the larger that cost. The worst case occurs of course when the information matrix is singular (see Section~\ref{sec:sing}), which leads to asymptotic local powers equal to the nominal level $\alpha$; more precisely, this situation entails that the best possible test is the trivial test, that is, the test discarding the observations and rejecting the null of reflective symmetry at level $\alpha$ whenever an auxiliary Bernoulli variable with parameter $\alpha$ takes value one.

Now, in order to take into account the aforementioned cost of not knowing~$\theta$, one can replace the central sequence $\Delta^{(n)}_{k;2}(\theta)$ with the so-called \emph{efficient central sequence}
\begin{eqnarray}
\Delta^{(n)eff}_{f_0,k;2}(\theta)&:=&\Delta^{(n)}_{k;2}(\theta)-\frac{\Gamma_{f_0,k;12}}{\Gamma_{f_0,k;11}}\Delta^{(n)}_{f_0,k;1}(\theta)\nonumber\\
&=&n^{-1/2}\sum_{i=1}^n\left(\sin(k(X_i-\theta))-\frac{\Gamma_{f_0,k;12}}{\Gamma_{f_0,k;11}}\varphi_{f_0}(X_i-\theta)\right).\label{eff}
\end{eqnarray}
This efficient central sequence can be seen as the orthogonal projection of $\Delta^{(n)}_{k;2}(\theta)$ onto the subspace orthogonal to $\Delta^{(n)}_{f_0,k;1}(\theta)$, which ensures that $\Delta^{(n)eff}_{f_0,k;2}(\theta)$ and $\Delta^{(n)}_{f_0,k;1}(\theta)$ are asymptotically uncorrelated.
An asymptotic test can then be easily obtained by considering a studentized version of $\Delta^{(n)eff}_{f_0,k;2}(\hat\theta)$. Unfortunately, by doing so, it can be shown that the effect of replacing $\theta$ with $\hat\theta$ is annihilated under $f_0$ only (the asymptotic linearity fails to hold for $g_0\neq f_0$). Therefore, rather than having as in the present paper a test which is valid under any density $f_0 \in \mathcal{F}$ with a fixed location $\theta$, we obtain a test which is valid under $f_0$ only (complete parametric test) and any value of $\theta$. Constructing tests which are completely distribution-free (with respect to both the underlying base density and the location parameter) is an ongoing research project.

\vspace{8mm}

\noindent ACKNOWLEDGEMENTS \vspace{0.5cm}

\noindent Christophe Ley thanks the Fonds National de la Recherche Scientifique, Communaut\'e fran\c caise de Belgique, for support via a Mandat de Charg\'e de Recherche.

\appendix

\section{Proof of Theorem~\ref{ULAN}}
Our proof relies on Lemma 1 of Swensen~(1985)---more precisely, on its extension in Garel and Hallin~(1995). The sufficient conditions for ULAN in those results readily follow from standard arguments (hence are left to the reader), once it is shown that $(\theta,\lambda)'\mapsto (f^k_{\theta,\lambda})^{1/2}(x)$ (see~(\ref{densitygen})) is quadratic mean differentiable at any $(\theta,0)'$, which we establish in the following lemma.
\vspace{1mm}

\begin{lem} \label{QMD}
Let $f_0\in\mathcal{F}$ and $k\in\N_0$, and assume that Assumption (A) holds. Define
$$
D_{\theta}(f^k_{\theta,0})^{1/2}(x):=-\frac{1}{2}\frac{\dot{f_0}(x-\theta)}{f_0^{1/2}(x-\theta)},
$$
and
$$
D_{\lambda}(f^k_{\theta,\lambda})^{1/2}(x)|_{\lambda=0}:=\frac{1}{2}f_0^{1/2}(x-\theta)\sin(k(x-\theta)).
$$
Then, for any $\theta\in[-\pi,\pi]$, we have that, as $(t,\ell)\to(0,0)$,
\begin{itemize}
\item[(i)] $\int_{-\pi}^\pi\left((f^k_{\theta+t,0})^{1/2}(x)-(f^k_{\theta,0})^{1/2}-tD_{\theta}(f^k_{\theta,0})^{1/2}(x)\right)^2dx=o(t^2)$,

\item[(ii)] $\int_{-\pi}^\pi\left((f^k_{\theta+t,\ell})^{1/2}(x)-(f^k_{\theta+t,0})^{1/2}-\ell D_{\lambda}(f^k_{\theta+t,\lambda})^{1/2}(x)|_{\lambda=0}\right)^2dx=o(\ell^2)$,

\item[(iii)]
$\int_{-\pi}^\pi\left(D_{\lambda}(f^k_{\theta+t,\lambda})^{1/2}(x)|_{\lambda=0}-D_{\lambda}(f^k_{\theta,\lambda})^{1/2}(x)|_{\lambda=0}\right)^2dx=o(1)$,

\item[(iv)]$\int_{-\pi}^\pi \left\{(f^k_{\theta+t,\ell})^{1/2}(x)-(f^k_{\theta,0})^{1/2}(x)-
\left(
\begin{array}{c}
t\\
\ell
\end{array}
\right)^{\prime}
\left(\begin{array}{c}D_{\theta}(f^k_{\theta,0})^{1/2}(x)\\
D_{\lambda}(f^k_{\theta,\lambda})^{1/2}(x)|_{\lambda=0}
\end{array}\right)\right\}^2
dx=o(||(t,\ell)'||^2).$
\end{itemize}
\end{lem}
\vspace{2mm}

\noindent \textbf{Proof of Lemma~\ref{QMD}.}   (i) By definition of $f^k_{\theta,0}$ we can rewrite the left-hand side of~(i) under the simpler form
\begin{equation}\label{half}
\int_{-\pi}^\pi \left(f_0^{1/2}(x-(\theta+t))-f_0^{1/2}(x-\theta)+\frac{1}{2}t\frac{\dot{f_0}(x-\theta)}{f_0^{1/2}(x-\theta)}\right)^2dx.
\end{equation}
Next, the a.e.-differentiability of $f_0$ (Assumption (A)) combined with the mean value theorem turns the latter expression into
\begin{eqnarray}
&&\int_{-\pi}^\pi \left(\frac{1}{2}t\frac{\dot{f_0}(x-\theta^*)}{f_0^{1/2}(x-\theta^*)}-\frac{1}{2}t\frac{\dot{f_0}(x-\theta)}{f_0^{1/2}(x-\theta)}\right)^2dx\nonumber\\
&&=\frac{1}{4}t^2\int_{-\pi}^\pi \left(\frac{\dot{f_0}(x-\theta^*)}{f_0^{1/2}(x-\theta^*)}-\frac{\dot{f_0}(x-\theta)}{f_0^{1/2}(x-\theta)}\right)^2dx\label{qmdexp}
\end{eqnarray}
with $\theta^*\in(\theta,\theta+t)$. Assumption~(A) ensures that $\frac{\dot{f_0}(x)}{f_0^{1/2}(x)}$ is continuous over the compact $[-\pi,\pi]$, hence its square can be bounded by a sufficiently large constant; consequently, the Lebesgue dominated convergence theorem implies that (\ref{qmdexp}) is $o(t^2)$.

(ii) Similarly, the left-hand side integral in (ii) can be re-expressed as
$$\int_{-\pi}^\pi f_0(x-\theta-t)\left((1+\ell\sin(k(x-\theta-t)))^{1/2}-1-\ell\frac{1}{2}\sin(k(x-\theta-t))\right)^2dx.$$
Exactly as for (i), the differentiability of $\sin(kx)$ allows us to re-write this integral under the form
$$
\frac{1}{4}\ell^2\int_{-\pi}^\pi f_0(x-\theta-t)\sin^2(k(x-\theta-t))\left(\frac{1}{(1+\ell^*\sin(k(x-\theta-t)))^{1/2}}-1\right)^2dx
$$
with $\ell^*\in(0,\ell)$. Since $\sin^2(kx)f_0(x)$ is integrable and $(1+\ell^*\sin(k(x-\theta-t)))^{-1}$ is bounded by a constant not depending on $\ell$ (indeed, we can take $\ell^*\leq\ell<1/2$ as $\ell\rightarrow0$, hence $1+\ell^*\sin(k(x-\theta-t))\geq1/2$ over $[-\pi,\pi]$ which does not depend on $\ell$), Lebesgue's dominated convergence theorem applies and yields the desired $o(\ell^2)$ quantity.

(iii) The left-hand side in (iii) equals
\begin{equation}\label{iii}
\frac{1}{4}\int_{-\pi}^\pi\left(f_0^{1/2}(x-(\theta+t))\sin(k(x-(\theta+t)))-f_0^{1/2}(x-\theta)\sin(k(x-\theta))\right)^2dx.
\end{equation}
Since $f_0^{1/2}(x)\sin(kx)$ is square-integrable, the quadratic mean continuity entails that (\ref{iii}) tends to zero as $t\rightarrow0$, hence is an $o(1)$ quantity.
 
(iv) The left-hand side in (iv) is bounded by $C(S_1+S_2+\ell^2S_3)$, where
$$S_1=\int_{-\pi}^\pi\left((f^k_{\theta+t,0})^{1/2}(x)-(f^k_{\theta,0})^{1/2}-tD_{\theta}(f^k_{\theta,0})^{1/2}(x)\right)^2dx,$$

$$S_2=\int_{-\pi}^\pi\left((f^k_{\theta+t,\ell})^{1/2}(x)-(f^k_{\theta+t,0})^{1/2}-\ell D_{\lambda}(f^k_{\theta+t,\lambda})^{1/2}(x)|_{\lambda=0}\right)^2dx,$$
and
$$S_3=\int_{-\pi}^\pi\left(D_{\lambda}(f^k_{\theta+t,\lambda})^{1/2}(x)|_{\lambda=0}-D_{\lambda}(f^k_{\theta,\lambda})^{1/2}(x)|_{\lambda=0}\right)^2dx.$$
The result then follows from (i), (ii) and (iii). \hfill $\square$

\section{Proof of Theorem~\ref{theorstudent1}.}

\noindent \textbf{Proofs of Theorem \ref{theorstudent1}.}
Fix  $f_0\in\mathcal{F}$. Part (i) of the theorem trivially follows from the Central Limit Theorem combined with the fact that
\begin{equation}\label{contig}
Q^{*(n);\theta}_{k}=\frac{|n^{-1/2}\sum_{i=1}^n\sin(k(X_i-\theta))|}{(n^{-1}\sum_{i=1}^n\sin^2(k(X_i-\theta)))^{1/2}}=\frac{|n^{-1/2}\sum_{i=1}^n\sin(k(X_i-\theta))|}{(\Gamma_{f_0,k;22})^{1/2}}+o_{\rm P}(1)
\end{equation}
as $n\rightarrow\infty$ under ${\rm P}^{(n)}_{(\theta,0)\pr;f_0}$. Regarding Part (ii), the situation is slightly more subtle but can also be readily solved by having recourse to the so-called ``Third Lemma of Le Cam'' (see Le Cam~1986). Under ${\rm P}^{(n)}_{(\theta,n^{-1/2}\tau\n_2)\pr;f_0,k'}$, the asymptotic normality of $\Delta^{(n)}_{k;2}(\theta)$ with mean $C_{f_0}(k,k')\tau_2$ and variance $\Gamma_{f_0,k;22}$ is obtained by establishing the joint normality of $\Delta^{(n)}_{k;2}(\theta)$ and \linebreak $\log\big(d{\rm P}^{(n)} _{(\theta, n^{-1/2}\tau_2^{(n)})\pr;f_0,k'}/ d{\rm P}^{(n)} _{(\theta,0)\pr;f_0}\big)$ under~${\rm P}\n_{(\theta,0)\pr;f_0}$ and then applying Le Cam's third Lemma (which holds thanks to the ULAN property). Part (ii) follows immediately since~(\ref{contig}) also holds under ${\rm P}^{(n)}_{(\theta,n^{-1/2}\tau\n_2)\pr;f_0,k'}$ by contiguity. Finally, Part (iii) trivially follows from (\ref{contig}) and the optimality features of the parametric test $\phi^{(n);\theta}_{f_0;k}$ for all $f_0\in\mathcal{F}$. \hfill$\square$
\end{document}